\documentclass[preprint]{aastex}	

\newcommand{\ergs}{ergs~s$^{-1}~$}
\newcommand{\mdot}{$\dot{{M}}$ } 

\received{}
\accepted{}
\lefthead{Bloser et al.}
\righthead{Spectral variation in 4U 1915-05 with {\em RXTE}}

\begin{document}

\title{{\em RXTE} Studies of X-ray Spectral Variations
with Accretion Rate in 4U 1915-05}
\author{P. F. Bloser, J. E. Grindlay}
\affil{Harvard-Smithsonian Center for Astrophysics, \\
60 Garden Street, Cambridge, MA 02138 \\
pbloser@cfa.harvard.edu}
\authoremail{pbloser@cfa.harvard.edu}
\and
\author{D. Barret, L. Boirin}
\affil{Centre d'Etude Spatiale des Rayonnements, CNRS/UPS,\\
9 Avenue du Colonel Roche, 31028 Toulouse Cedex 04, France\\
Didier.Barret@cesr.fr}


\begin{abstract}

We present the results of detailed spectral studies of the 
ultra-compact low mass X-ray binary (LMXB) 4U 1915-05 carried out 
with the {\em Rossi X-ray Timing Explorer (RXTE)} during 1996.  4U
1915-05 is an X-ray burster (XRB) known to exhibit a $\sim 199$-day
modulation in its 2--12 keV flux.  Observations were performed with the
PCA and HEXTE instruments on {\em RXTE} at roughly one-month 
intervals to sample this long-term period and study 
accretion rate-related spectral changes.  We obtain good fits with a
model consisting of a blackbody and an exponentially cut-off power
law.  The spectral parameters are strongly correlated with both the
broad-band (2--50 keV)  
luminosity and the position in the color-color diagram, with the
source moving from a low hard state to a high soft state as the
accretion rate increases.  
The blackbody component appears to drive the spectral evolution.
Our results are consistent with a geometry in which
the soft component arises from an optically thick boundary layer and
the hard component from an extended Comptonizing corona.  Comparing
our results with those of a similar study of the brighter source 4U 1820-30
(Bloser et al. 2000), we find that the two ultra-compact LMXBs occupy
similar 
spectral states even though the transitions occur at very different
total luminosities.  

\end{abstract}

\section{Introduction}
\label{sec-intro}

Recent advances in broad-band X-ray astronomy observations have led
to rapid progress in the study of Low Mass X-ray Binaries
(LMXBs) containing low-magnetic field neutron stars (NSs).  Many of
these
systems exhibit bright X-ray bursts at low accretion rates; those that
do are known as X-ray bursters (XRBs).  Narrow-band observations at high
energies by BATSE and SIGMA have shown that XRBs can emit hard X-rays
at $\sim 100$ keV (Barret \& Vedrenne 1994, Tavani \&
Barret 1997 and references therein).  The lack of simultaneous low
energy observations has until recently made it difficult to relate the
hard X-ray emission 
to the softer blackbody and disk blackbody components long known to
exist in XRBs and other NS systems (e.g. Mitsuda et al. 1984).  The
broad-band spectral sensitivities of the {\em Rossi X-ray Timing
Explorer} ({\em RXTE}, Bradt, Rothschild, \& Swank 1993) and {\em
BeppoSAX} (Boella et al. 1997) have now allowed detailed
modeling of the complete X-ray to hard X-ray (1--200 keV) spectra of
several XRBs with a variety of complex physical models including
blackbody, disk blackbody, power law, Comptonization, reflection, and
line components (e.g., Church et
al. 1998; Guainazzi et al. 1998; in 't Zand et al. 1999; Olive et
al. 1999, Piraino et al. 1999a; Piraino et al. 1999b; Barret et
al. 1999, 2000; Bloser et al. 2000).  These observations promise to
reveal the relationship between the hard and soft spectral components,
the geometry of the emitting regions, and the dependence of each on
the mass accretion rate.   

In this paper we present the spectral results of {\em RXTE}
observations of a 
particularly interesting XRB, the 50-minute binary 4U 1915-05.  This
system is well-known as the first dipping source discovered (White \&
Swank 1982; Walter et al. 1982), and was the first XRB to show clear
evidence for a binary nature (Walter et al. 1982).  The source has
generated extensive 
interest due to the fact that its dip period, $\sim 50$ minutes, is
close to but significantly shorter than the period of modulation of
its optical counterpart, a $m_V \sim 21$ blue star (Grindlay et
al. 1988; Callanan et al. 1995).  This has led to numerous
interpretations, including the possibility that the system is a
hierarchical triple (Grindlay et al. 1988).  From 
evolutionary calculations on low-mass systems with short orbital
periods, Nelson, Rappaport \& Joss (1986) find that the companion is
likely a hydrogen-deficient star that is neither fully degenerate nor
burning helium.  Smale et al. (1988) argue that the lack of X-ray
eclipses, combined with the orbital parameters expected for such a
companion star, constrain the orbital inclination to be $< 79^{\circ}$.
From analysis of
an X-ray burst that showed photospheric radius expansion, Yoshida
(1993) derived a distance for 4U 1915-05 of 9.3 kpc.
A long-term X-ray modulation
with a rough period of 199 days is also seen in {\em Vela 5B} data
(Priedhorsky \& Terrell 
1984a), indicating a long-term modulation in the accretion rate.

The goal of the present investigation is to observe the broad-band
spectrum of the persistent (non-burst and non-dip) emission from 4U
1915-05 with the PCA and HEXTE instruments on {\em RXTE} in order to
study the relationships between spectral parameters and accretion
rate.  (Analysis of these same data for the 50-minute X-ray dip period
and related periodic behavior is presented by Chou, Bloser, \&
Grindlay 2000.)
The 199-day variation in the flux from 4U 1915-05 is similar
to that in another compact XRB, 4U 1820-30, which has a 176-day
period (Priedhorsky
\& Terrell 1984b).  4U 1820-30 only exhibits X-ray bursts in the
lowest state, indicating that the modulation is due to accretion rate
variations and not, for example, to disk precession effects. 
Although 4U 1915-05 is a dimmer source and thus
able to emit X-ray bursts over a wider range of its luminosity, this
similarity implies that the 199-day modulation is also due to
accretion rate variations.  Thus these two sources provide good
laboratories for the study of accretion-related spectral changes.  
We have observed 4U 1915-05 at roughly monthly intervals over the
course of a year to sample this 199-day period.
(A re-analysis of the {\em Vela 5B} data by Smale \& Lochner (1992)
puts the significance of this period at only $\sim 80$\%; however, the
source may still be counted on to vary its luminosity on this
rough timescale.)
A similar analysis of 4U 1820-30 has been presented by Bloser et
al. (2000).   

\section{Previous Spectral Observations of 4U 1915-05}
\label{sec-prev}

Previous spectral observations of 4U 1915-05 have been performed by a
variety of experiments.  In early work, simple, one-component
models were fit to the spectra.  White \& Swank (1982) fit a power law
(PL) with
photon index $\alpha =$ 1.6--1.7 to {\em OSO-8} data and a thermal
bremsstrahlung model with $kT = 12$ keV to (non-simultaneous) {\em
Einstein}  MPC+SSS 
data.  The {\em Einstein}
luminosity was 1.2--1.5$ \times 10^{37}(D/10 {\rm kpc})^2$ \ergs
(0.5--60 keV).  Smale et al. (1988) fit {\em EXOSAT} data with a PL
with $\alpha = 1.8$ and found a luminosity of $7.6 \times
10^{36}(D/10 {\rm kpc})^2$ \ergs (0.1--15 keV).  Extensive
observations were performed by {\em Ginga} (Yoshida 1993), which
obtained data over a wide range of source luminosities.  These spectra
could be fit by a combination of a cut-off power law (CPL; an approximation
of Comptonization) and a broad gaussian at 7 keV.  Using the {\em
Ginga} results of Yoshida (1993), Barret et al. (1996) defined three
distinct 
intensity/spectral states for 4U 1915-05 with typical values of the
power law index $\alpha$, exponential cut-off energy $E_c$, and 1--37 keV
luminosity $L$: a high/very soft state ($\alpha = 1.0$, $E_c \sim $ 8--10
keV, $L = 1.2 \times 10^{37}$ \ergs); an 
intermediate/hard state ($\alpha = 1.7$, $E_c \gtrsim $ 50
keV, $L =$ 6--7 $\times 10^{36}$ \ergs); and a low/soft state ($\alpha
= 1.5$, $E_c \sim $ 25
keV, $L = 3 \times 10^{36}$ \ergs).  Barret et al. (1996) note that
the evidence for softening at low luminosity is marginal and needs to
be confirmed.  

Two high-energy experiments on {\em CGRO} failed to detect 4U 1915-05
above 50 keV. 
Bloser et al. (1996) analyzed four years of BATSE data and report a
$2\sigma$ upper limit of $6.3 \times 10^{36}$ \ergs on the 20--100 keV
luminosity of the source for 10 day integrations.  Barret et
al. (1996) did not detect 4U 1915-05 in a pointed OSSE observation,
reporting $2\sigma$ upper limits in two energy bands of $1.7 \times
10^{35}$ \ergs (50--98 keV) and $1.3 \times 10^{35}$ \ergs (98--158
keV).  

More recent observations have allowed two-component models to
be fit.  Church et al. (1997) fit 0.7--10 keV {\em ASCA} data with a
combination of a PL and blackbody (BB), finding a best-fit
blackbody temperature $kT_{BB} = 2.14$ keV and $\alpha = 2.42$.  (Note
however that Ko et al. (1999) are able to fit the same data with a
single power 
law with $\alpha \sim 1.75$; the blackbody of Church et al. 1997
shares flux with the power law near 10 keV, and so their power law is
steeper.)  Morley 
et al. (1999) use the same model 
on {\em ROSAT} data (0.1--2 keV) and find $kT_{BB} = 1.95$ keV and $\alpha
= 2.32$.  Finally, the first true hard X-ray detection of 4U 1915-05
was reported by Church et al. (1998), who detect the source above 100
keV in a broad-band (0.1--100 keV) {\em BeppoSAX} observation.  They
fit a combination of a CPL and BB and find $kT_{BB} = 1.62$ keV,
$\alpha = 1.61$, and $E_c = 80 \pm 10$ keV.  

A brief summary of previously-reported spectral results on 4U 1915-05 is
presented in Table~\ref{tab-prev}.

\section{Observations and Analysis}
\label{sec-obsanal}

\subsection{{\em RXTE} Observations and Data Reduction}

The PCA instrument on {\em RXTE} is made up of five Xenon proportional
counter units (PCUs, numbered 0--4) sensitive from 2--60 keV with a
total area of about 6500 
cm$^2$ (Bradt, Rothschild, \& Swank 1993).  
The HEXTE instrument consists of two clusters of four NaI(Tl)/CsI(Na)
phoswich scintillation detectors sensitive from 15--200 keV with an
effective area of 1600 cm$^2$; these clusters rock on and off source
to obtain background measurements (Rothschild et al. 1998).  The ASM
is comprised of three Scanning Shadow Cameras (SSCs) with
one-dimensional slit masks and $6^{\circ} \times 90^{\circ}$ fields of
view that monitor the fluxes of bright X-ray sources several times a
day in the 2--12 keV band (Levine et al. 1996).  

Nine joint PCA/HEXTE pointed observations of 4U 1915-05 were
performed at roughly monthly intervals between 1996 February 10 and
1996 October 29.  An additional ten observations were performed on
consecutive days between 1996 May 14 and 23.  A complete log of the
observations is given by Boirin et al. (2000).  The {\em RXTE}/ASM
2--12 keV light curve of 4U 1915-05 during this 
period is shown in Figure~\ref{fig-asm}, with the times of our pointed
observations marked.  Each point represents the one-day average of the
ASM count rate.  Although the 199-day period reported by Priedhorsky
\& Terrell (1984a) is not obvious in Figure~\ref{fig-asm}, in
agreement with the findings of Smale \& Lochner (1992), it is clear
that we have observed the source in a variety of intensity states
ranging from $\sim$ 0.2--2 ASM counts s$^{-1}$, or $\sim$ 2.7--27
mCrab. 

The data were divided into 70 segments (typical length $\sim 1200$
seconds) of persistent emission, 
excluding bursts and dips, and reduced using the
standard {\em RXTE} analysis tools contained in FTOOLS 4.2.
For the HEXTE spectral analysis we used standard archive mode data and
extracted spectra with {\em saextrct} v4.0b.  Deadtime corrections were 
computed using {\em hxtdead} v0.0.1.
For our color-color diagram and 
PCA spectral analysis we used the ``Standard 2'' data, which provide
128 energy channels between 2 and 100 keV with 16-second time
resolution.  Spectra were extracted with {\em saextrct}, and the
v2.2.1 response matrices (January 1998) were used.

Complications arose in the analysis of the PCA data from the fact that
the PCA gain was lowered twice during 1996.
Our first two observations, performed on 1996 February 10 and March
13, occurred during PCA Gain Epoch 1, while the remaining observations
occurred during Gain Epoch 3.  Different background models and response
matrices are needed for each Gain Epoch.  During both Epoch 1 observations
the source was ``bright'' (defined as having a count rate greater than
40 cts s$^{-1}$ PCU$^{-1}$), and so we used the ``new sky-VLE''
(August 1999) models recommended for Epoch 1 to generate a background
using {\em pcabackest} v2.1b.  We found
that the 
background for the March 13 observation was computed correctly, but
that the February 10 background was badly overestimated.  We attempted
to correct
the February 10 background by renormalizing the activation component,
as recommended by the Guest Observer Facility web pages. 
When producing the color-color diagram (see Section~\ref{sec-ccd}),
however, we found that the points from these data 
segments were badly misplaced.  We therefore excluded the
February 10 data from further analysis.  The Epoch 3 backgrounds were
computed using {\em pcabackest} with the appropriate Epoch 3 bright
and faint source background models, depending on the source count rate (see
Table 1 in Boirin et al. (2000)).

In this
paper we present a spectral 
analysis of the persistent emission.  A timing analysis of the
persistent emission and bursts, including the discovery of twin
kilohertz quasi-periodic oscillations (QPOs), is given by Boirin et
al. (2000).  An 
analysis of the X-ray dip period and optical period is given by Chou,
Bloser, \& Grindlay (2000), and a study 
of the spectral evolution in dipping will be the subject of a forthcoming
paper.  

\subsection{Color-Color Diagram and Parameterization of Accretion
Rate}
\label{sec-ccd}

A color-color diagram (CCD) was produced from the PCA data with one
point for each of the 70 data segments.  Only PCUs 0, 1, and 2 were on
continuously during all observations, and so only data from these PCUs are
included here.
The soft and hard colors are defined as the ratios of
background-subtracted PCA count rates in the bands 
3.5--6.4 keV and 2.0--3.5 keV, and 9.7--16.0 keV and 6.4--9.7 keV,
respectively.  Data from Gain Epoch 1 (the 1996 March 13 observation)
are difficult to include in the same diagram because the
channel-to-energy conversion is different, leading to differences
in the energy band channel boundaries.  We have attempted to
divide the March 13 counts into the proper energy bands by employing
fractional channel boundaries and assigning the correct fraction of
counts in the ``in-between'' channels to either the higher or lower band
based on the fitted spectral shape.  
The CCD is shown in Figure~\ref{fig-ccd}.  We observe
4U 1915-05 in the lower and upper banana
branches, two of the states associated with ``atoll''
sources (Hasinger \& van der Klis 1989).  Although it appears in
Figure~\ref{fig-ccd} that the points with the highest hard color form
a separated cluster, suggesting that
4U 1915-05 enters the island state as well,
the timing analysis shows that the source never exhibits the high
frequency noise typically observed in atoll sources in the island
state (Boirin et al. 2000).  We shall refer to the lowest state in our
data as the ``{\em RXTE} low state,'' and note that it refers not to
the island state but the lower banana branch.
The Epoch 1 data segments
are indicated by the open squares.  As these data represent the source
at its brightest,  we expect them to lie in the upper banana portion
of the CCD, and indeed this is where they appear.  Due to the
background and gain uncertainties mentioned above, however, their exact
position should be treated with caution.

It is generally believed that in atoll sources the mass accretion rate
\mdot increases monotonically as the source moves from the island
state, through the lower banana, and into the upper banana branch
(Hasinger \& van der Klis 1989).  M\'endez et al. (1999) defined the
parameter $S_a$ to measure the position of an atoll source within its
CCD and thus parameterize its accretion rate.  We follow M\'endez et
al. (1999; see also Bloser et al. 2000) and approximate the shape of
the CCD track with a spline, 
as shown in Figure~\ref{fig-ccd}.  
$S_a$ is then defined as the
distance along this curve, as measured from the {\em RXTE} low state.  
Thus we may use $S_a$ in addition to the source luminosity to study
accretion rate-related changes in spectral parameters.  We have not
used the suspect Epoch 1 data points in generating the $S_a$ track.
In order to use $S_a$ as the basis of
comparison for all of our data, we have assigned 
values of $S_a$ to the Epoch 1 points according to the
ratio of their
total luminosity to luminosities of the Epoch 3 data.

\subsection{Spectral Fitting}
\label{sec-fitting}

For our spectral analysis data segments were grouped according to
their position in the CCD.  Spectra were first extracted for each
of the 70 data segments.  These spectra were then grouped into eleven
equal bins in $S_a$ and combined together, with an 
average value of $S_a$ and the total exposure time computed for each
bin.  Thus we had a total of eleven spectra to fit (see
Table~\ref{tab-fits1}).  

The spectral analysis was performed using XSPEC v10.0 (Arnaud 1996).
Spectra for 
each PCU and HEXTE cluster were reduced separately and combined within
XSPEC.  Due to systematic uncertainties in the response matrices, only
HEXTE data above 20 keV were used.  Based on fits to archival Crab
data (Bloser et al. 2000; see also Sobczak et al. 1999; Barret et
al. 2000; Wilms et al. 1999) we use only PCA data between 2.5 and
20.0 keV from PCUs 0 \& 1.  (Although PCU 4 also gives good fits for
Crab data, it was rarely on during our 4U 1915-05 observations.)  We
note that Epoch 1 Crab data gave larger systematic deviations in the
residuals than Epoch 3 data, especially below 5--6 keV.  The
relative normalizations of the two PCUs and two HEXTE clusters were left as
free parameters, since there still exist large uncertainties in the
relative flux normalizations; all spectral model normalizations reported
here are obtained from PCU 0.  
A systematic error of 1\% was added to all PCA channels using {\em
grppha}.  In Figure~\ref{fig-twospec} we show two 
representative PCA and HEXTE spectra of 4U 1915-05, one from the {\em
RXTE} low
state and one from the high state.  The hardening of the
spectrum at low luminosity is clear.  In the {\em RXTE} low state the
source is 
detected by HEXTE only up to 50 keV, as opposed to the high energy
detection by {\em BeppoSAX} (Church et al. 1998).

All spectra were initially fit with the two-component model used by
Church et al. (1998) for the {\em BeppoSAX} data consisting of a
cut-off power law plus a blackbody (CPL + BB).  This model provided a
good fit to all eleven spectra, and in all cases the inclusion of the BB
improved the fit greatly over the CPL alone.  In addition, the
residuals of the fit to the brightest spectrum (1996 March 13) showed
an excess near 6 keV which could be fit with a gaussian line at
$6.14^{+0.18}_{-1.07}$ keV with $\sigma = 0.23^{+1.91}_{-0.23}$ keV.
Including this gaussian improved the fit significantly ($> 99.9$\%
significance from an F-test), and the line energy is roughly consistent
with that of a neutral Fe $K\alpha$ line, though the parameters are highly
uncertain.  We note that the March 13 data are from Epoch 1, and
so effects in the residuals near 5--6 keV should be treated with
caution.  In Figure~\ref{fig-fit} we
show two examples of spectra from PCUs 0 and 1 and HEXTE clusters A
and B fit with the CPL + BB model.  These
are the same two spectra shown in Figure~\ref{fig-twospec}; now we
show the raw counts with the folded model, the residuals, and the
unfolded spectra with each component plotted separately.  The gaussian
is included in the March 13 spectrum.

Additional models were fit to the data as well.  The spectra of
several  XRBs have recently been fit with detailed Comptonization
models (Barret et al. 1999; Barret et al. 2000; Guainazzi et
al. 1998; Piraino et al. 1999a; Piraino et al. 1999b; Bloser et
al. 2000).  We therefore fit our 4U 1915-05 spectra with the
recently-developed Comptonization model of Titarchuk (1994), called 
the {\em CompTT} model in XSPEC.  
The parameters of interest are the temperature
$kT_e$ and optical depth $\tau$ of the Comptonizing electron cloud,
and the temperature of the cool seed photons
$kT_W$, assumed to follow a Wien law.
Good fits were found to all but the
dimmest three observations with a combination of this model and a
blackbody (CompTT + BB).  In the case of the dimmest observations no
cutoff could be determined, and so the electron temperature $kT_e$
could not be constrained.  Both the CPL and CompTT models were also
tried with the blackbody component replaced by a multi-color disk
blackbody (DBB; Mitsuda et al. 1984).  Here the parameters are 
the color temperature of the
inner disk $kT_{in}$ and the projected inner disk radius
$R_{in}\sqrt{\cos \theta}$, where $\theta$ is the inclination of the
system.  Physically realistic fits could only be found with this model
for the brightest six observations.

The insensitivity of the PCA below 2 keV did not allow us to
determine the hydrogen column density $N_H$ from our spectral fits.
It was therefore necessary to freeze $N_H$ at previously-determined
values for all spectra.  The values of $N_H$ found from previous X-ray
observations are listed in Table~\ref{tab-prev}.  The hydrogen column
density measured from the galactic H {\sc i} survey of Stark et
al. (1992) is $N_H = 1.96 \times 10^{21}$ cm$^{-2}$ (Smale et
al. 1988; preliminary Stark results reported as private
communication).  Based on 
these numbers, which are all in reasonable 
agreement, we took a rough average and set $N_H = 2.0 \times 10^{21}$ 
cm$^{-2}$ and kept it frozen for all our spectral fits.

\section{Results}
\label{sec-res}

The best-fit spectral parameters for the CPL + BB model are given in
Table~\ref{tab-fits1} with the average value of $S_a$ for all included
data.  The parameters for the CompTT + BB model
are given in Table~\ref{tab-fits2} for the eight spectra for which
they could be derived.  In Table~\ref{tab-fits1} we include the total
integration time for each spectrum, the effective radius of the
BB emitting surface $R_{BB}$, and the  ratio $L_{BB}/L_H$ of
the BB luminosity to the hard spectral component luminosity in the
2--50 keV band covered by the PCA and HEXTE.  In
Table~\ref{tab-fits2} we give the best-fit temperature of the seed
photons $kT_W$, the Comptonizing $y$-parameter, defined as $y =
4kT_e\tau^2/m_ec^2$, and the effective Wien radius $R_W$ of the seed
photons (in 't Zand et al. 1999; Bloser et al. 2000).  The fits are
very good for both models for all spectra; in fact, the values of
$\chi^2_{\nu}$ are always $< 1$, indicating that the systematic errors
may have been over-estimated.

In Figure~\ref{fig-params1} we show the spectral
parameters of the CPL + BB model as a function of
$S_a$, and thus presumably of \mdot.  It is immediately obvious that
all parameters are strongly correlated with $S_a$.  As $S_a$ increases
the blackbody temperature $kT_{BB}$ increases steadily and the
relative contribution of the blackbody component increases
from $< 15$\% to $\sim 30$\%.  The power law photon index $\alpha$
decreases smoothly as $S_a$ increases, while the cutoff energy $E_c$
makes a rather sudden transition at $S_a \sim 1.3$ from $> 100$ keV to
$\sim 40$ keV.  (Note however that even the lower limits on $E_c$ for
the dimmest three spectra lie well above the highest energy, 50 keV,
at which the source is detected, and so should be treated with
caution.)  This transition value of $S_a$ corresponds roughly to the
``vertex'' at the leftmost corner of the CCD.  From the behavior of
the spectral parameters it is clear 
that 4U 1915-05, like most systems, moves from a low, hard state to a
high, soft state as 
the accretion rate increases.  Note that when fitting a CPL model it
is not  
$\alpha$ but $E_c$ that indicates the hardness of the spectrum.  Note
also that the BB component appears to be the dominant factor in
driving the spectral evolution of the source: $kT_{BB}$ is very
strongly correlated with $S_a$, and the large increase in $L_{BB}/L_H$
corresponds to the probable quenching of the Comptonizing corona
indicated by the drop in $E_c$.

In Figure~\ref{fig-params2} we show the spectral
parameters of the CompTT + BB model as a function of
$S_a$ for the brightest eight spectra.  Again the parameters are
correlated with $S_a$, though not as strongly (due partly to the
absence of the three lower state spectra).  There is an indication that
$\tau$ increases from $\sim 6$ to $\sim 14$ and $kT_e$ decreases from
$\sim 7$ to $\sim 4$ as $S_a$ increases, but the error bars are
large.  The BB parameters are similar to those in the CPL + BB model,
though again with larger uncertainties.  Although the derived
CompTT + BB
parameters are similar to those found in other sources 
(Barret et al. 1999; Barret et al. 2000; Guainazzi et
al. 1998; Piraino et al. 1999a; Piraino et al. 1999b; Bloser et
al. 2000), the smaller error bars and more robust correlations found
in Figure~\ref{fig-params1} indicate that the CPL + BB model provides
a better description of the 4U 1915-05 data.

Fits were also performed using CPL + DBB and CompTT + DBB models.  Only
the brightest six observations gave acceptable fits.  The CPL and
CompTT parameters were similar to those found using the CPL + BB and
CompTT + BB models.  The DBB parameters in both cases did not show any
correlation with $S_a$.  For the CPL + DBB model the parameters were
$kT_{in} \sim 0.6$ keV and $R_{in} \sim 8$ km.  For the CompTT + DBB
model the parameters were $kT_{in} \sim 0.9$ keV and $R_{in} \sim 6$
km, with greater scatter and large error bars.

\section{Discussion}
\label{sec-disc}

Our {\em RXTE} results for the CPL + BB model may be compared to those of
previous investigations.  Based on the spectral parameters of the CPL
+ BB model shown in Figure~\ref{fig-params1}, we have observed 4U
1915-05 in the intermediate/hard and high/very soft states defined by
Barret et al. (1996) from the {\em Ginga} data (Yoshida 1993).  The
transition occurs at a broad-band luminosity of $\sim $7--8$ \times
10^{36}$ \ergs, consistent with the {\em Ginga} results.  We do not
see evidence for a low/soft state in the {\em RXTE} data; however, our
observations do not reach as low a luminosity as the {\em Ginga}
observations.  Translating our lowest 
luminosity value into the 1--37 keV {\em Ginga} band gives $\sim 6 \times
10^{36}$ \ergs, well above the value of $\sim 3 \times 10^{36}$ \ergs
for the {\em Ginga} low/soft state.  In addition, the {\em BeppoSAX}
observation described by Church et al. (1998) found a well-determined
cut-off energy of $80 \pm 10$ keV, 
which is excluded at the 90\% confidence level for our
{\em RXTE} low state.  This could indicate spectral softening at low
luminosity.  
Extrapolating their spectral parameters 
into the 2--50 keV band of {\em RXTE} gives a luminosity of $4.5 \times
10^{36}$ \ergs, also lower than the lowest value seen by {\em RXTE}.
Thus we cannot rule out a low/soft state at luminosities fainter than
those observed by {\em RXTE}.  Such a low/soft state would be
difficult to explain in terms of the current physical picture of XRBs.

We can extrapolate our spectral results to higher energy for
comparison with the non-detections by BATSE and OSSE (Bloser et
al. 1996; Barret et al. 1996).  Our {\em RXTE} low state spectrum would
give a 20-100 
keV luminosity of $3.8 \times 10^{36}$ \ergs.  This is consistent with the
$2\sigma$ upper limit of $6.3 \times 10^{36}$ \ergs found by BATSE for
the 20-100 keV band (Bloser et al. 1996).  In the 50--98 keV band our
{\em RXTE} low state 
spectrum gives a luminosity of $1.2 \times 10^{36}$ \ergs, which is
well above the OSSE upper limit of $1.7 \times 10^{35}$ \ergs for this
band.  Our high state spectrum, however, gives a consistent 50--98 keV
value of 
$4 \times 10^{34}$ \ergs.  Thus our data support the conclusion of
Barret et al. (1996) that 4U 1915-05 was in the high state during the
OSSE observation.

One of the major issues in the interpretation of XRB spectra is the
origin of the hard versus soft spectral components (e.g., White, Stella, \&
Parmar 1988; Guainazzi et al. 1998; Olive et al. 1999, Barret et
al. 1999, 2000).   While it seems clear from spectral and timing
similarities with black hole systems that the hard component must
arise from Comptonization in some form of hot corona (Barret et
al. 2000), both the source 
of the cool photons that are up-scattered and the source of the soft
thermal component are the subject of debate.  The soft spectral
component could be dominated by thermal emission from either the inner
accretion disk or an optically thin boundary layer where the accreted
matter hits the NS surface.  (It seems likely that in reality soft
emission could come from both sources, but current instruments are
unable to separate the two.)  Our good CPL + BB fits, which employ a
single-temperature blackbody appropriate for an optically thick
boundary layer, support the latter of these two pictures.  The
temperature $kT_{BB}$ increases monotonically with $S_a$, as might be
expected if 
the BB luminosity increases monotonically with \mdot,
while the small derived equivalent radius $R_{BB}$ implies that only a
narrow equatorial band around the NS is emitting.  Such small values
of $R_{BB}$ have been seen before in 4U 1915-05 (Church et al. 1997;
Morley et al. 1999; Church et al. 1998), and in other
sources as well (White, Stella, \& 
Parmar 1988; Church et al. 1998; Bloser et al. 2000; Church and
Baluci\'nska-Church 2000).   
In this picture the hot corona is then supported by soft photons from
the disk (hard emission powered by the boundary layer is not favored
due to the similarities between NS and BH spectra).
A complication is the fact that the
ratio of the BB flux to the hard PL flux $L_{BB}/L_H$ is always less
than 30\%.  Sunyaev \& Shakura (1986) showed that, if
relativistic effects are taken into account, the luminosity of the
boundary layer should be at least equal to that of the disk.  Again,
many authors have noted a similar discrepancy in other sources (White,
Stella, \& Parmar 1988; Barret et al. 1999, 2000).  In the case of
4U 1915-05 a possible explanation is simply that, at such a high
inclination, part of the boundary layer emission is being blocked by
the inner disk.  The decrease in $R_{BB}$ as $S_a$ increases
(Figure~\ref{fig-params1}) may support this notion, since perhaps the
inner disk can become thicker as the accretion rate increases and
obscure more of the boundary layer.

In contrast, our attempts to fit the soft component with a disk
blackbody model produced unconvincing results, due to the small values
of the inner disk radius derived.
The value of $R_{in}$
must be adjusted for the effects of electron scattering in the inner
disk according to the relation $R_{eff} = f^2R_{in}$ where $f = 1.7$
(Ebisawa et al. 1994; Shimura \& Takahara 1995; Barret et al. 2000).
Assuming the maximum 
inclination angle of $79^{\circ}$ (Smale et al. 1988) this gives inner
disk radii of $\sim 53$ km for the CPL + DBB model and $\sim 40$ km
for the CompTT + DBB model.  These values are only slightly larger
than the radius of the marginally stable orbit around a 1.4
$M_{\odot}$ NS, and it seems unlikely that in a dim source like 4U
1915-05 the disk could extend this far in.  In addition, the value of
$R_{in}$ shows no correlation with $S_a$.  Boirin et al. (2000) show
that the frequency of the kilohertz QPOs
in 4U 1915-05 increases with $S_a$ for all observations, which is
usually interpreted in terms of the inner disk radius decreasing (e.g.,
Miller, Lamb, \& Psaltis 1998).  Thus the inner disk in 4U
1915-05 is 
probably moving inward as the accretion rate increases, a fact not
reflected in the derived values of $R_{in}$.  We conclude that the DBB
model is not a good description of the soft component in our data.  It
is possible that the edge-on nature of 4U 1915-05 is making emission from
the inner accretion disk difficult to detect directly.

Our data are thus more consistent with a picture in which the soft
spectral component is produced by an optically
thick boundary layer which is partially blocked from view by the inner
disk.  This soft component is the most strongly correlated with the
accretion rate, as seen in Figure~\ref{fig-params1}, and is therefore
the component which drives the spectral changes, despite the fact that
the hard component contains a larger fraction of the flux.
As the temperature $kT_{BB}$ increases, soft thermal photons
quench the Compton up-scattering in the corona, and the cut-off energy
$E_c$ of the power law decreases.  It appears in
Figure~\ref{fig-params1} that the sharp decrease in $E_c$ coincides
roughly with an increase in the relative luminosity of the BB.  This
geometrical picture
is supported by the work of Church et al. (1998), Morley et
al. (1999), and Church et al. (1997), all of whom find that the dips
in 4U 1915-05 may be modeled by a compact blackbody emitter being
occulted while 
an extended hard X-ray emitter is only partially covered.  A spectral
analysis of the PCA dip data will be the subject of a future paper.
We note, however, that Bloser et al. (2000) found that the soft spectral
component in 4U 1820-30, a source very similar to 4U 1915-05 in many
ways (though with a lower inclination), {\em could} be described as a
DBB.  In this picture the hard 
power law emission arises from a hot, optically thin flow {\em
within} the inner disk radius (Barret et al. 2000; Bloser et
al. 2000).  Since 4U 1915-05 is a dimmer source than 4U 1820-30, at
first glance it would seem to be an even better candidate to support a
hot, optically thin accretion flow.

In Figure~\ref{fig-comp} we show our
CPL + BB fits for 4U 1915-05  with those of Bloser et al. (2000) for
4U 1820-30, now showing spectral parameters as a function of total
2--50 keV luminosity to allow direct comparison.  The assumed
distances are 9.3 kpc for 4U 1915-05 and 6.4 kpc for 4U 1820-30.  It
is clear that 4U 1820-30 is always a more luminous source than 4U
1915-05, and yet the behavior of the CPL component is qualitatively
similar.  Using $E_c$ as an indicator, both sources move from a low,
hard state to a high, soft state as the luminosity increases, and in
both sources the transition happens relatively suddenly.  The
transition occurs at different luminosities in each source, however:
7--8$ \times 10^{36}$ \ergs in 4U 1915-05, $3 \times 10^{37}$ \ergs in
4U 1820-30.  The value of $E_c$ is higher in 4U 1915-05 than in 4U
1820-30 in equivalent states, as might be expected in a lower
luminosity source, 
but they do not form a continuous 
sequence with luminosity; $E_c = 9$ keV at $1.44 \times 10^{37}$
\ergs in 4U 1915-05, but $E_c = 23$ keV at $2.31 \times 10^{37}$
\ergs in 4U 1820-30.  Thus something other than total luminosity, or
accretion rate, must determine when an XRB makes the transition to a
low state.  The behavior of the BB component does appear to form a
continuous sequence with luminosity: $kT_{BB}$ increases steadily with
luminosity in 4U 1915-05 until it reaches 2.5 keV at $1.4 \times
10^{37}$ \ergs.  In 4U 1820-30, the luminosity is always above $1.4 \times
10^{37}$ \ergs, and $kT_{BB}$ is nearly constant at $\sim 2.5$ keV. 

We note that Church and Baluci\'nska-Church (2000) have fit a sample
of LMXBs containing NS primaries with a CPL + BB model and found that,
as the luminosity increases, $R_{BB}$ increases considerably while
$kT_{BB}$ decreases slightly.  Our results for 4U 1915-05 and 4U
1820-30 (Bloser et al. 2000) seem to contradict this trend, as the
values of $kT_{BB}$ are roughly the same for the sources while
$R_{BB}$ increases only very slightly for 4U 1820-30, despite a large
difference in luminosity (Figure~\ref{fig-comp}).  The Church and
Baluci\'nska-Church (2000) sample covers a much larger range of
luminosity (up to several $\times 10^{38}$ \ergs), however, and
differences between 4U 1915-05 and 4U 1820-30 could well be lost in
the scatter.  More compelling is the fact that each of the sources in
the Church and Baluci\'nska-Church (2000) sample were observed only
once.  We have followed 4U 1915-05 throughout its CCD, and $kT_{BB}$
clearly increases with increasing accretion rate and luminosity.
Again, it seems that the total luminosity of a source does not
determine its state so much as its own internal variations.  In
addition, the Church and Baluci\'nska-Church (2000) sources were
observed only between 1--10 keV, perhaps limiting the accuracy of the
spectral fits.

The differences in spectral parameters between the two sources are
reflected in their CCDs.  In Figure~\ref{fig-ccds} we compare the CCDs
of 4U 1915-05 and 4U 1820-30 (Figure 2 of Bloser et al. 2000).  The
$S_a$ tracks for both sources are indicated for clarity.  Although both
sources cover the same range in soft color, 4U 1820-30 nearly
always has a lower value of the hard color, reflecting its low
values of $E_c$.  The two CCDs overlap at the 4U 1820-30 island state,
and these points fall in the same
region of the CCD as do points in the fourth $S_a$ bin of 4U 1915-05
($S_a = 1.36$).
The values of $E_c$ at this location are indeed similar, within the
errors:  $31.25^{+19.16}_{- 7.48}$ keV for 4U 1915-05, $23.07^{+
2.91}_{- 2.36}$ keV for 4U 1820-30.  The 4U 1915-05 power law is
harder, however ($\alpha = 1.85$ versus 2.05), and the blackbody
temperature is lower ($kT_{BB} = 1.51$ keV versus 2.57 keV). 
The 2--50 keV luminosities are $5.3 \times 10^{36}$ \ergs for 4U
1915-05 and $2.3 \times 10^{37}$ \ergs for 4U 1820-30, a factor of
$\sim 4$ difference.  Although
it is difficult to disentangle the effects of spectral shape,
inclination, and neutral absorption, Figures~\ref{fig-comp} and
\ref{fig-ccds} suggest that different sources can occupy similar
states and regions of the CCD at very different luminosities.

4U 1915-05 is also similar to 4U 1820-30 in that it exhibits kilohertz
QPOs whose behavior depends on position within the CCD (Boirin et
al. 2000).  As noted above, the frequency of the QPOs increases with
$S_a$, indicating that the inner disk radius is shrinking with increasing
accretion rate and that the area of the disk is therefore increasing.
Increased soft photon emission from the disk is doubtless 
contributing to the quenching of the hot corona seen in
Figure~\ref{fig-params1}.  The RMS amplitude of the QPOs decreases
with count rate (although, interestingly, a correlation with $S_a$ is
not apparent), and 
they are not detected at all in the upper banana branch.  If the
decreasing amplitude is indeed related to accretion rate, this is in
agreement with the sonic point beat frequency model of Miller, Lamb,
\& Psaltis (1998), in which the increasing optical depth of the
accretion flow at higher accretion rates inhibits the radiation drag
which creates the QPOs in the first place.  Twin kilohertz QPOs are
detected only in a narrow range of $S_a$, which corresponds to the
``vertex'' in the CCD of Figure~\ref{fig-ccd}.  This is also the range
in $S_a$ in which the source makes its rapid transition from the {\em
RXTE} low state (Figure~\ref{fig-params1}) to the high state.  In the
sonic point model, 
the second QPO is produced when a portion of the gas is funneled by
the NS magnetic field and produces hot spots that modulate the
radiation drag at the NS spin frequency.  There is no obvious relation
to the cutoff energy in this model; perhaps the inner disk reaches a
region where the magnetic field is strong enough to produce this
funneling at an $S_a$ that just happens also to produce a BB flux
strong enough to quench the corona.

\section{Conclusions}
\label{sec-conc}

We have observed 4U 1915-05 in the lower and upper banana states with
{\em RXTE} and find that they correspond to the low/intermediate and
high/very soft states defined by Barret et al. (1996) from {\em Ginga}
data (Yoshida 1993).  The CPL + BB model provides a good fit to the
data, and all spectral parameters show strong correlations with both
the 2--50 keV luminosity and the accretion rate parameterized by
$S_a$.  The BB temperature $kT_{BB}$ is the most strongly correlated
with $S_a$, and so this component is primarily responsible for driving
the accretion-related spectral changes.
Spectral parameters are not as well constrained using the
CompTT + BB model, and models involving a DBB component do not give
realistic fits.  Our results are consistent with a geometry in which
the soft component arises from an optically thick boundary layer and
the hard component from an extended Comptonizing corona.  4U 1915-05
was observed in similar states as was 4U 1820-30 (Bloser et al. 2000),
but the states correspond to different values of the total
luminosity in the two systems.  The CPL component is more prominent in
4U 1915-05 than in 4U 1820-30, as expected for a dimmer source, and
this is consistent with differences in the two sources' color-color
diagrams.

\acknowledgments
We wish to thank A. P. Smale and J. H. Swank for early discussions of
the data.  
This paper had made use of 
quick-look results provided by the ASM/{\em RXTE} team.  This work was
supported in part by NASA grant NAG5-3298.
PFB acknowledges partial support from
NASA GSRP grant NGT5-50020.


\clearpage

\figcaption[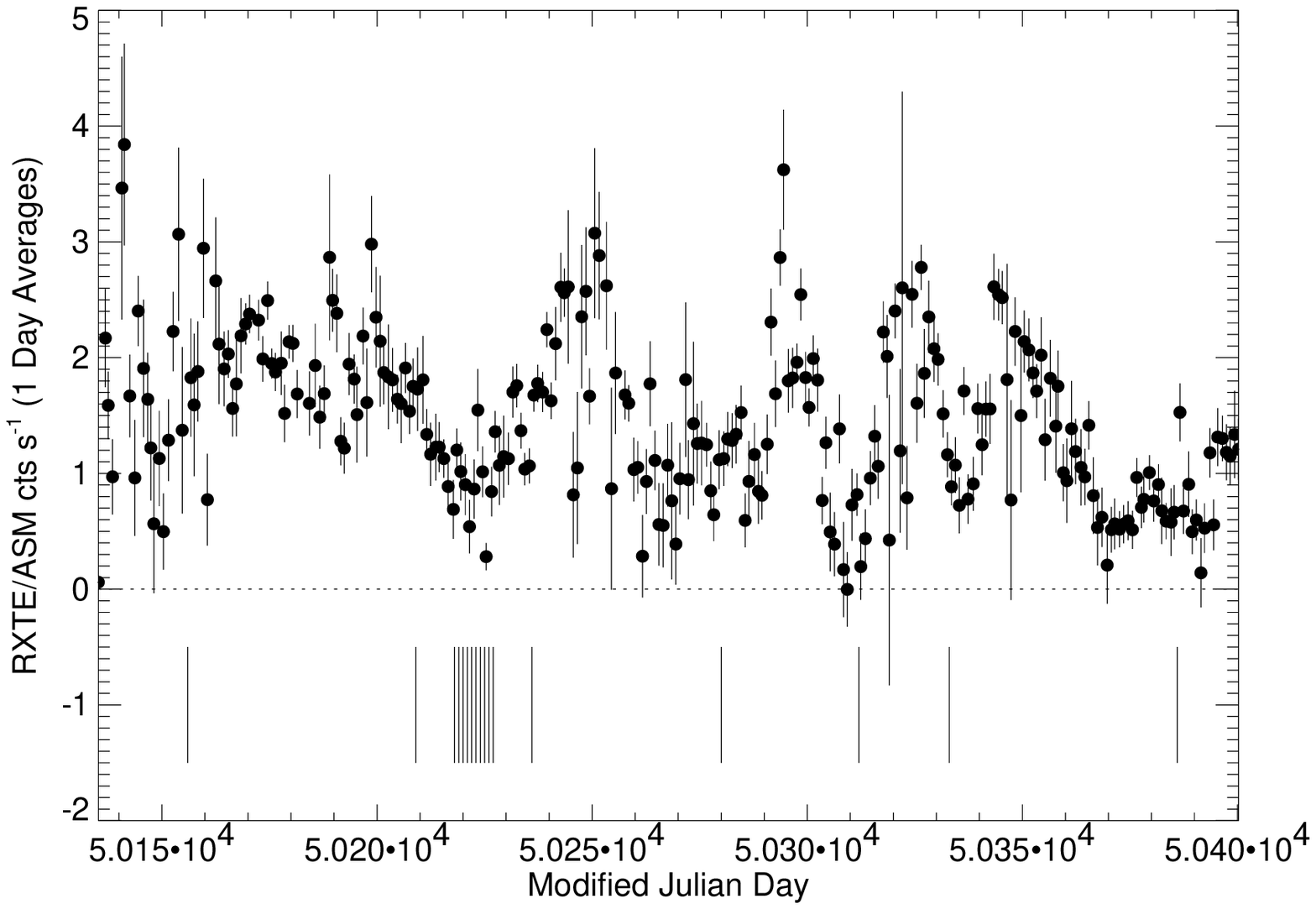]{{\em RXTE}/ASM light curve of 4U 1915-05
(2--12 keV) from 
1996 March -- October (MJD 50150 = 1996 March 7).  Each
point represents a 1-day average.  Pointed PCA/HEXTE
observation times are marked 
with vertical lines.  
\label{fig-asm}}

\figcaption[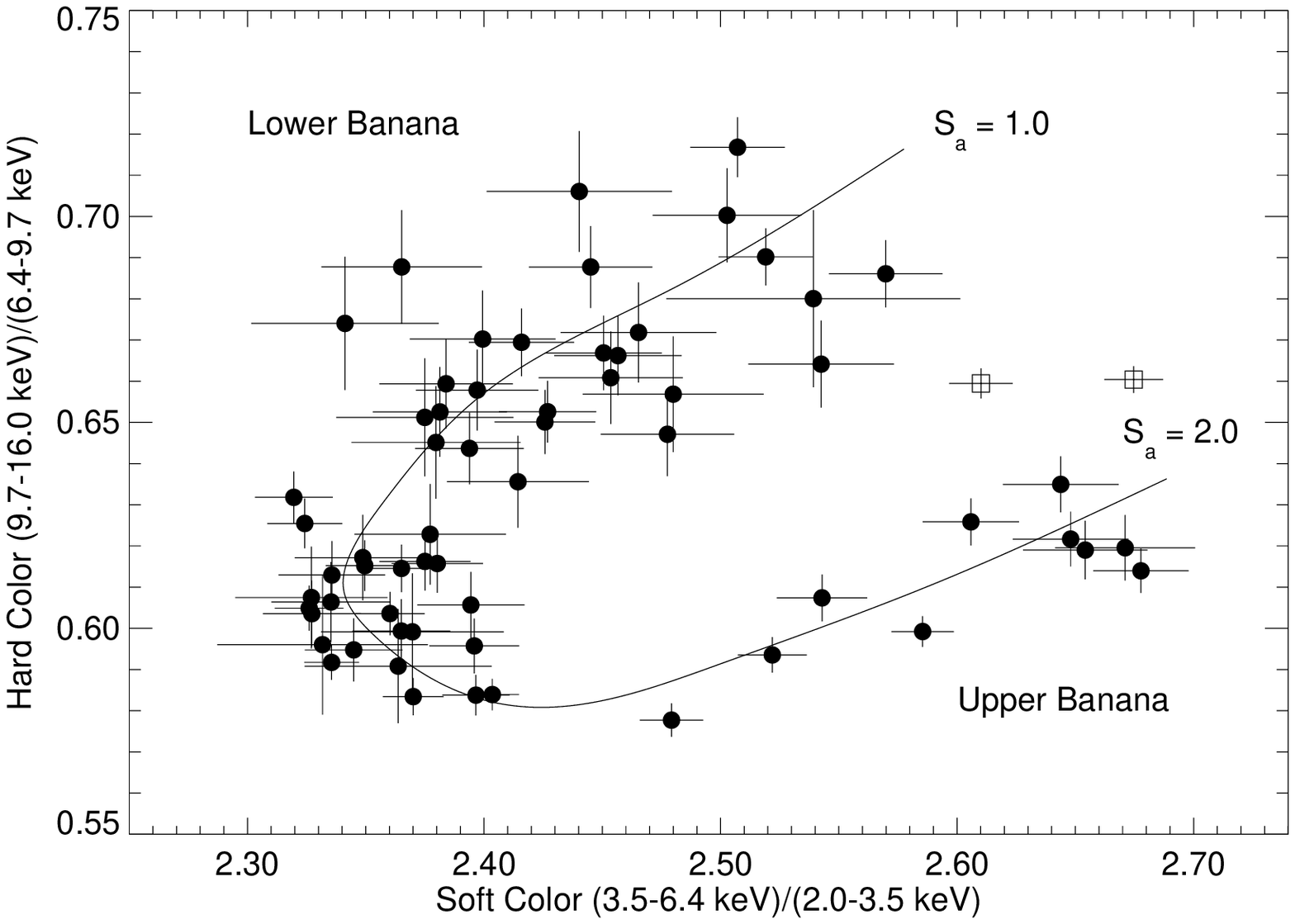]{Color-color diagram of 4U 1915-05.
Each point represents one 
of the 70 data segments from Boirin et al. (1999), which have typical lengths
of 1200 s.  The soft and hard colors are defined as the ratio of
background-subtracted count
rates in the bands 3.5--6.4 keV and 2.0--3.5 keV, and 9.7--16.0 keV
and 6.4--9.7 keV, respectively.  Filled symbols indicate data
from Epoch 3, open symbols data from Epoch 1.  The Epoch 1 points may
be positioned incorrectly, due to difficulties in matching
energy bands between the two gain epochs.
The position within the diagram is parameterized by the
variable $S_a$, the distance along the line.  
\label{fig-ccd}}

\figcaption[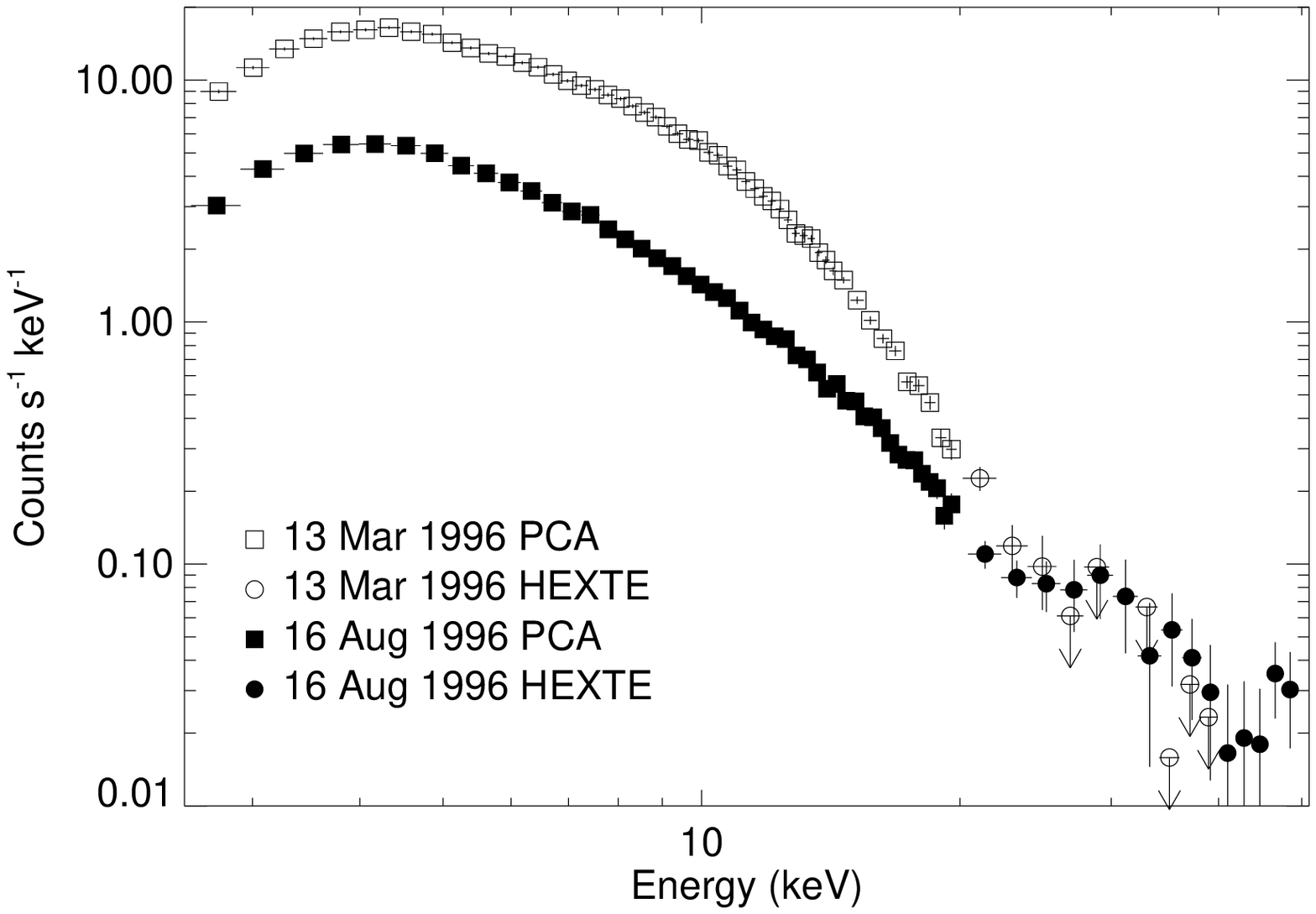]{Representative PCA (PCU 0) and HEXTE
(cluster B) spectra of 4U 1915-05.  On 1996
March 13 the source was in a high state.  
On 1996 August 16 the source was in the low
state and is detected with HEXTE up to 50 keV.  The hardening
of the spectrum with decreasing luminosity is clear.
\label{fig-twospec}}

\figcaption[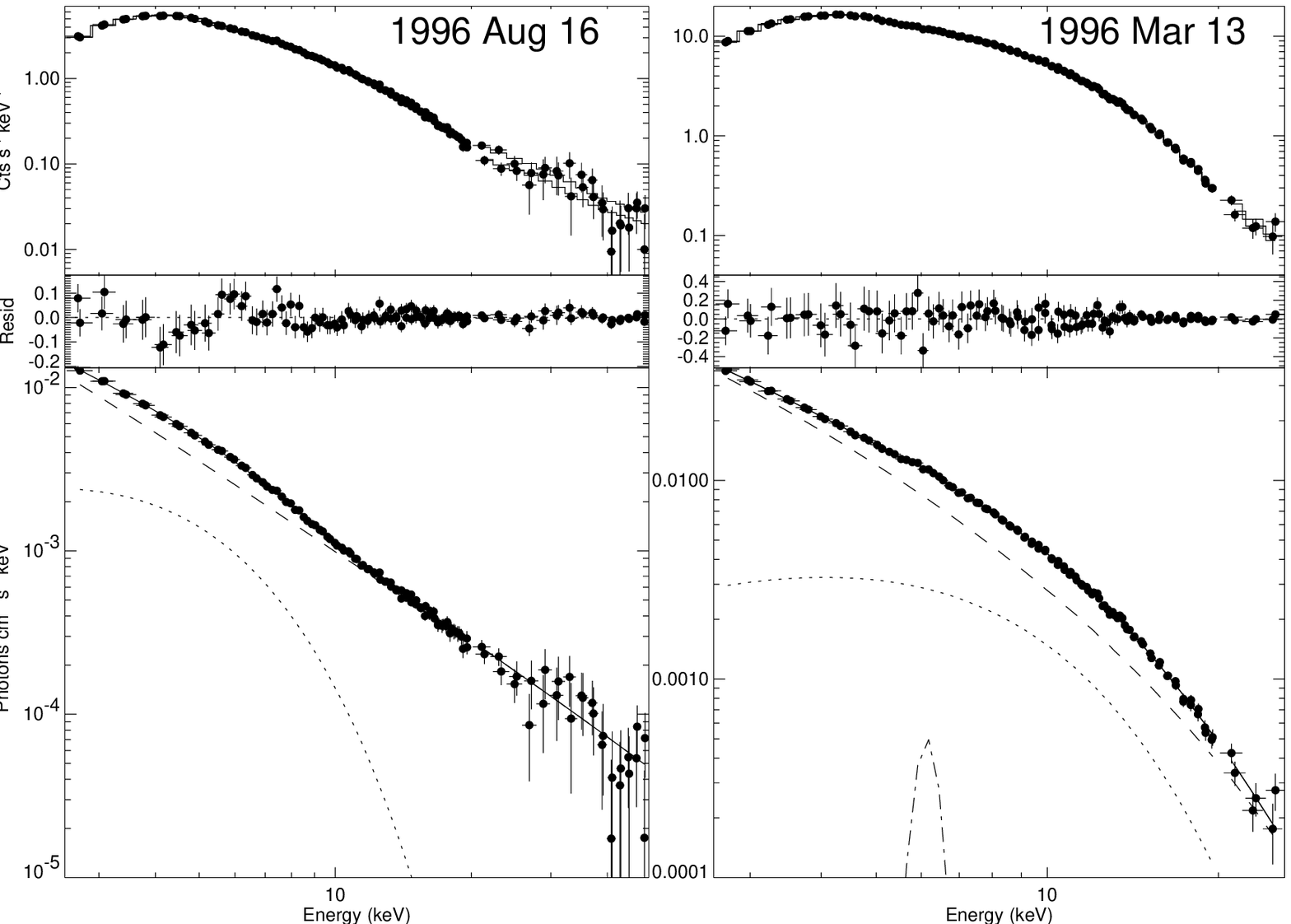]{Examples of 4U 1915-05 spectra fit with the
CPL + BB 
model.  The top panel shows the raw count rate spectrum with the fitted
model folded through the instruments' responses,  the middle panel shows
the residuals of the fit, and the bottom panel shows the unfolded
spectrum with the individual model components.  The dotted line is the
blackbody, the dashed line is the CPL model, and the dash-dot line
is the gaussian line (right panel only).  On the left is the low
state spectrum 
of 1996 August 16, and on the right is the high state spectrum of 1996
March 13. 
\label{fig-fit}}

\figcaption[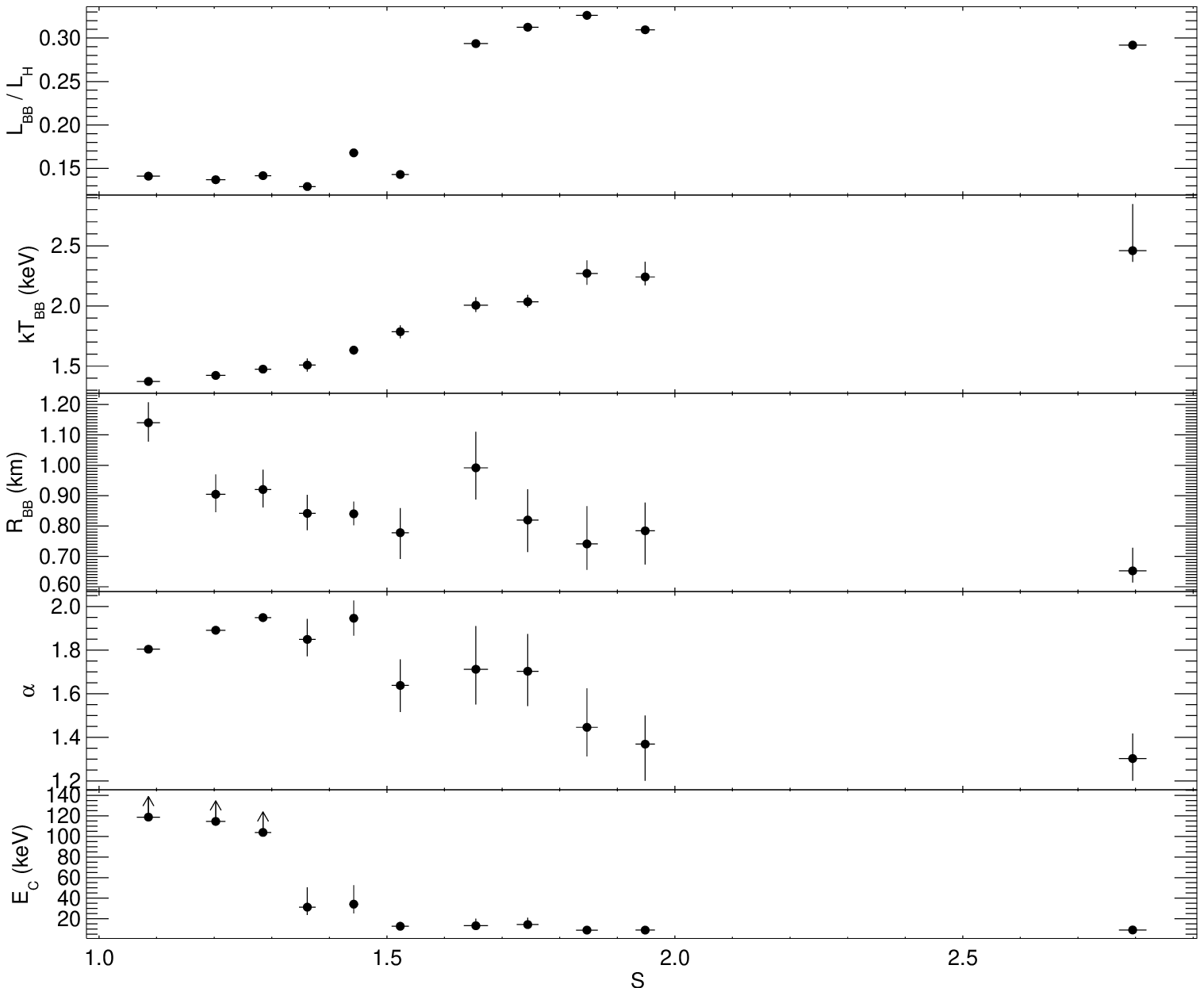]{Variation of spectral parameters
for the CPL + BB model
with accretion rate, 
parameterized by $S_a$.  Here $kT_{BB}$ = blackbody temperature,
$R_{BB}$ = blackbody radius, $\alpha$ = power law photon index, and $E_c$
= power law cut-off energy.  Error bars are $1\sigma$
for one interesting parameter.  The point with the highest $S_a$ is
Epoch 1 data, and this $S_a$ value has been rescaled relative to the
others according to total flux.
\label{fig-params1}}

\figcaption[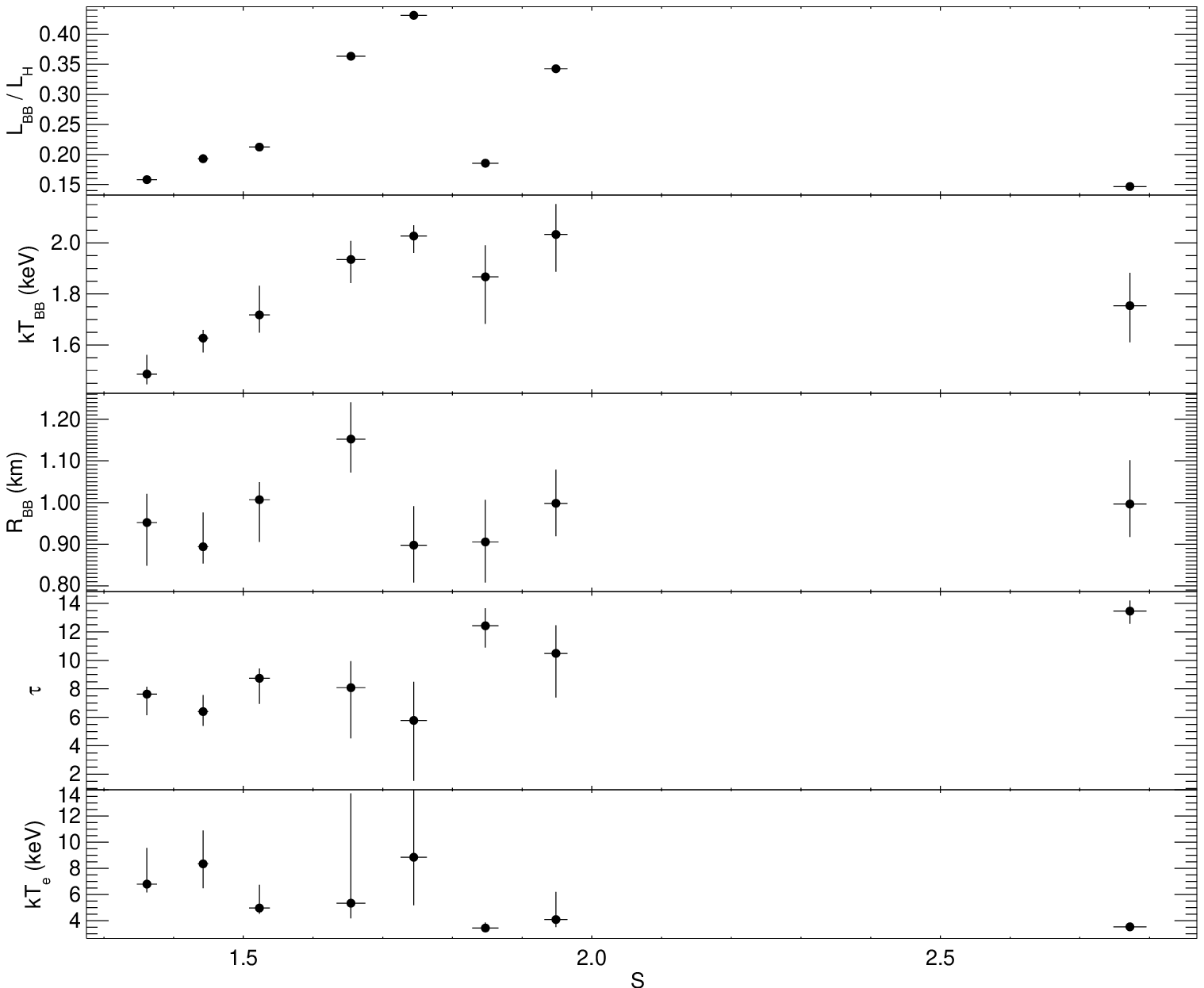]{Variation of spectral parameters
for the CompTT + BB model with accretion rate, 
parameterized by $S_a$.  Only the brightest eight spectra could be fit
with this model, as the faintest three lacked a discernible cutoff.
Here $\tau$ = Comptonization optical depth and
$kT_e$ = Comptonizing electron temperature. 
The point with the highest $S_a$ is 
Epoch 1 data, and this $S_a$ value has been rescaled relative to the
others according to total flux.
\label{fig-params2}}

\figcaption[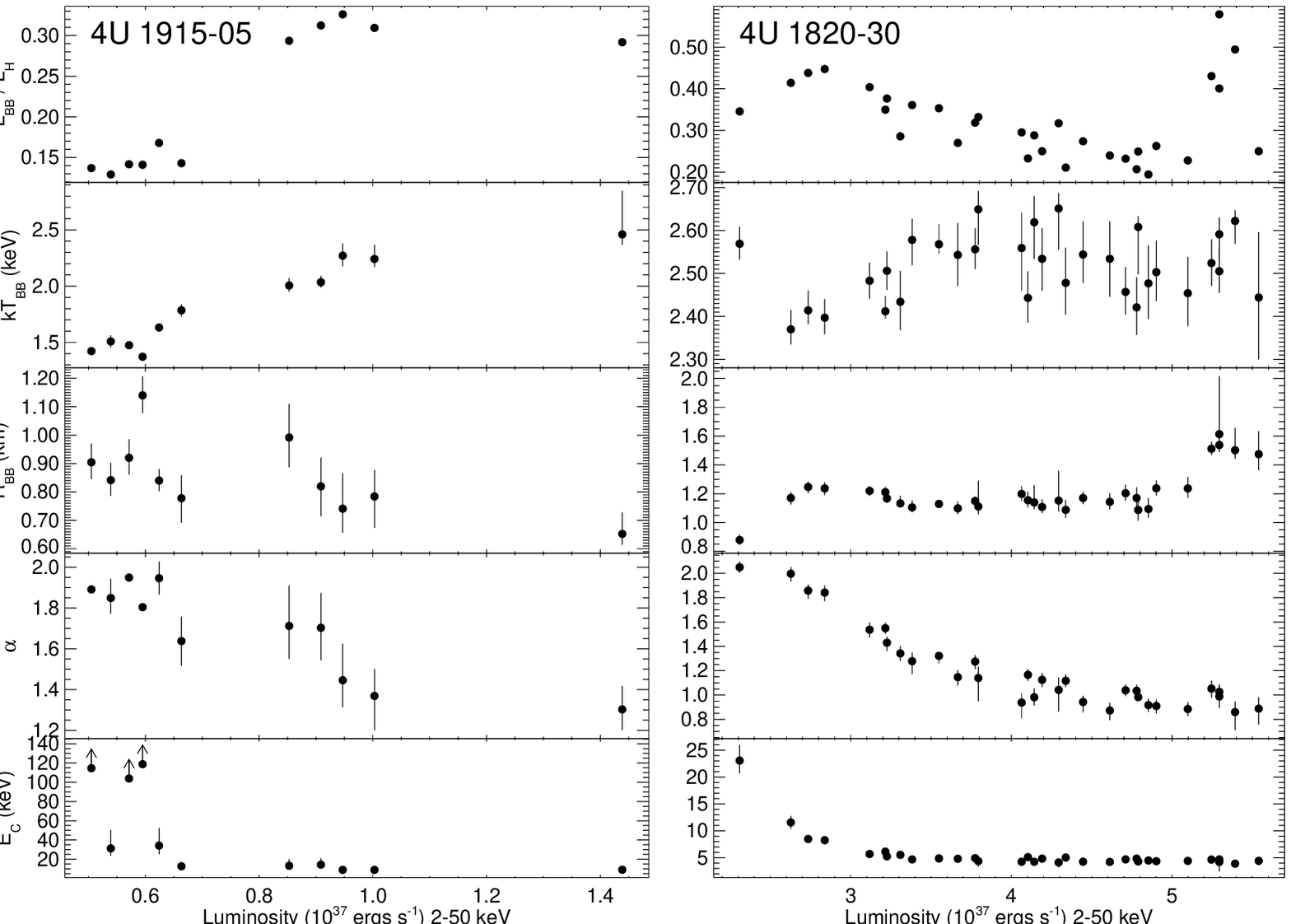]{Comparison of fitted parameters for the
CPL + BB model for 4U 1915-05 (this paper; assumed distance = 9.3 kpc)
and 4U 1820-30 (Bloser et 
al. 1999; assumed distance = 6.4 kpc).  The behavior of the power law
component is qualitatively similar but quantitatively different in the
two sources in 
that the rapid transition from a low hard state to a high soft state
takes place at very different total luminosities.  The behavior of the
blackbody component is more quantitatively consistent in the sense
that $kT_{BB}$ increases with luminosity until it reaches $\sim 2.6$
keV.  
\label{fig-comp}}

\figcaption[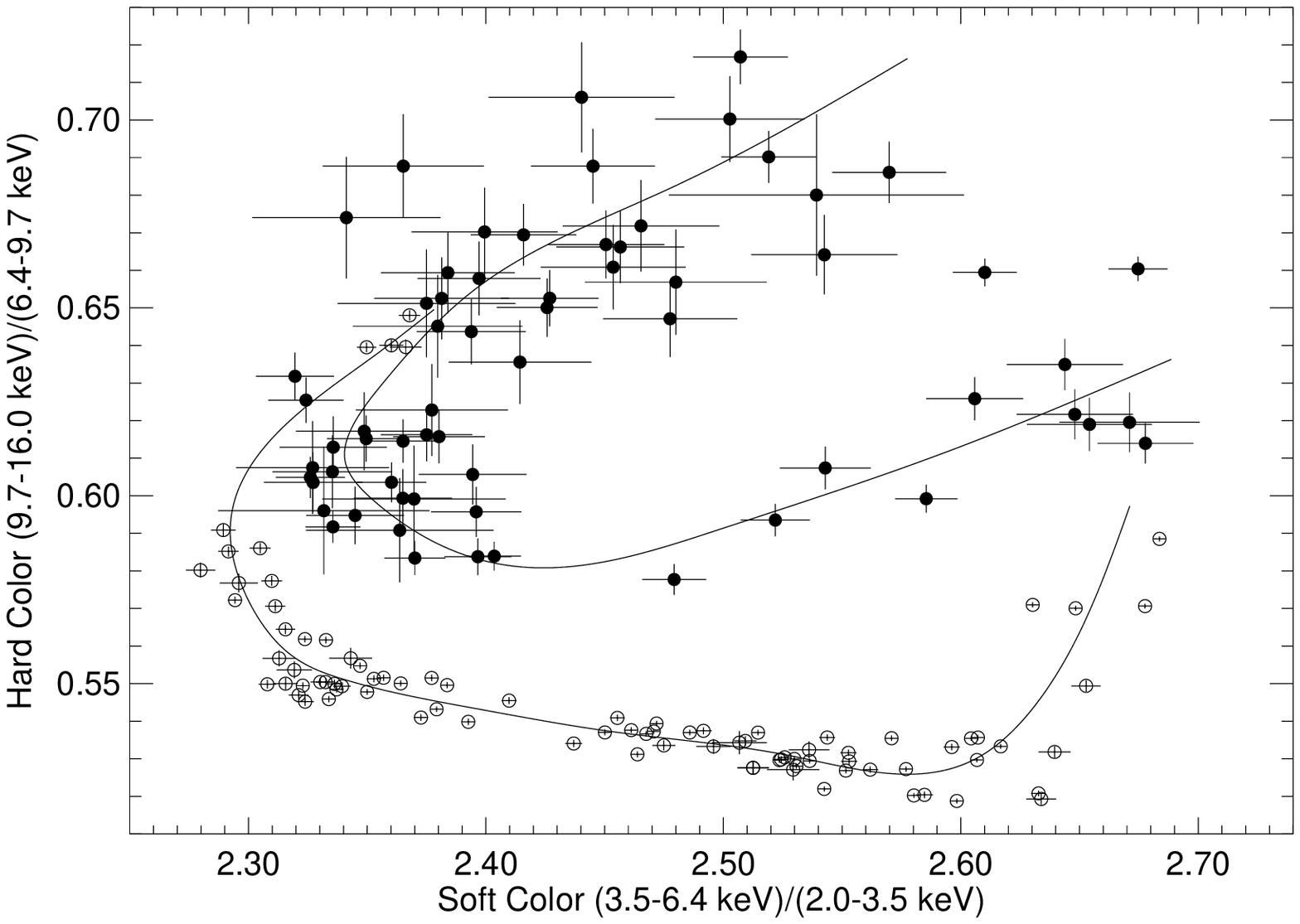]{Comparison of the color-color diagrams
of 4U 1915-05 (open symbols) and 4U 1820-30 (closed symbols; Bloser et
al. 1999).   The $S_a$ tracks are included for clarity.  Except for
the 4U 1820-30 island state, the hard color in 4U 1915-05 is always
greater than that in 4U 1820-30, though both sources cover the same
range in soft color.
\label{fig-ccds}}


\clearpage

\begin{deluxetable}{lclccc}
\footnotesize
\tablecaption{Previous Spectral Results on 4U 1915-05\label{tab-prev}} 
\tablewidth{0pt}
\tablehead{
\colhead{Experiment} &
\colhead{Year} &
\colhead{Model\tablenotemark{a}} &
\colhead{$N_H$} &
\colhead{Luminosity\tablenotemark{b}} &
\colhead{Ref} \\
\colhead{} &
\colhead{} &
\colhead{} &
\colhead{($10^{21}$ cm$^{-2}$)} &
\colhead{($10^{37}$ \ergs)} &
\colhead{} 
}
\startdata
{\em OSO-8} & 78 & PL ($\alpha =$ 1.6--1.7) & ... & ...  & 1\\
{\em Einstein} (MPC+SSS) & 79 & BR ($kT_{BR} = 12$ keV) & 3.0 & 1.0--1.3
(0.5--60 keV) & 1\\ 
{\em EXOSAT} & 83 \& 85 & PL ($\alpha = 1.8$) & 2.1 & 0.66 (0.1--15
keV) & 2\\ 
{\em Ginga} (high) & 90 & CPL ($\alpha = 1.0$, $E_c \sim $ 8--10
keV) & 1.3 & 1.2 (1--37 keV) & 3\\
{\em Ginga} (intermediate) & 87 \& 88 & CPL ($\alpha = 1.7$, $E_c
\gtrsim $ 50 keV) & 0.2 & 0.6--0.7 & 3\\
{\em Ginga} (low) & 88 & CPL ($\alpha = 1.5$, $E_c \sim $ 25 keV) & 0.6 &
0.3 & 3\\ 
{\em CGRO} (OSSE) & 95 & ... & ... & $< 0.017$ (50--98 keV) & 4\\
{\em CGRO} (BATSE) & 91--95 & ... & ... & $< 0.63$ (20--100 keV) & 5\\
{\em ASCA} & 93 & BB+PL ($kT_{BB} = 2.14$ keV,$\alpha = 2.42$) & 4.75
& 0.29 (1--10 keV) & 6\\
{\em ROSAT} (PSPC) & 92 & BB+PL ($kT_{BB} = 1.95$ keV, $\alpha =
2.32$) & 3.9 & ... (0.1--2 keV) & 7\\
{\em BeppoSAX} & 97 & BB+CPL ($kT_{BB} = 1.62$ keV, $\alpha = 1.61$,
 & 3.2 & 0.25 (1--10 keV) & 8\\ 
 & & \hspace{1.5cm}$E_c = 80 \pm 10$ keV) & & & \\
\enddata
\tablenotetext{a}{PL = Power Law; BR = Bremsstrahlung; CPL = Cut-off
Power Law; BB = Blackbody; $\alpha =$ power law photon index; $kT_{BR}
= $ Bremsstrahlung temperature; $kT_{BB} =$ blackbody temperature; $E_c
=$ exponential cut-off energy}
\tablenotetext{b}{Luminosity assumes a distance of 9.3 kpc}
\tablerefs{
(1) White \& Swank 1982; (2) Smale et al. 1988;
(3) Yoshida 1993;
(4) Barret et al. 1996; (5) Bloser et al. 1996; (6) Church et al. 1997;
(7) Morley et al. 1999; (8) Church et al. 1998
}
\end{deluxetable}


\clearpage

\begin{deluxetable}{lccccccccccc}
\tablecaption{Spectral Fits of 4U 1915-05 with the CPL + BB
Model\label{tab-fits1}} 
\tablewidth{0pt}
\tablehead{
\colhead{$S_a$} & 
\multicolumn{2}{c}{Int Time (s)} & 
\colhead{$kT_{BB}$\tablenotemark{a}} &
\colhead{$R_{BB}$\tablenotemark{b}} &
\colhead{$\alpha$\tablenotemark{c}} &
\colhead{$E_c$\tablenotemark{d}} &
\colhead{$L$\tablenotemark{e}} &
\colhead{$L_{BB}/L_H$\tablenotemark{f}} &
\colhead{$\chi^2_{\nu}$} \\
\colhead{} &
\colhead{PCA} &
\colhead{HEXTE} &
\colhead{(keV)} &
\colhead{(km)} &
\colhead{} & 
\colhead{(keV)} &
\colhead{} &
\colhead{} &
\colhead{}
}
\startdata
1.09 &  7755 & 2509 & $ 1.37^{+ 0.03}_{- 0.02}$ & $ 1.14^{+ 0.07}_{-
0.06}$ & $ 1.80^{+ 0.02}_{- 0.02}$ & $>119$ & 0.60 & 0.12 & 0.68 \\
1.20 &  8249 & 2669 & $ 1.42^{+ 0.03}_{- 0.03}$ & $ 0.90^{+ 0.07}_{-
0.06}$ & $ 1.89^{+ 0.02}_{- 0.02}$ & $>115$ & 0.51 & 0.12 & 0.79 \\
1.28 &  7361 & 2485 & $ 1.48^{+ 0.03}_{- 0.03}$ & $ 0.92^{+ 0.07}_{-
0.06}$ & $ 1.95^{+ 0.02}_{- 0.02}$ & $>104$ & 0.57 & 0.12 & 0.81 \\
1.36 & 14030 & 4630 & $ 1.51^{+ 0.06}_{- 0.05}$ & $ 0.84^{+ 0.06}_{- 0.06}$ & $ 1.85^{+ 0.09}_{- 0.08}$ & $ 31.25^{+19.16}_{- 7.48}$ & 0.54 & 0.11 & 0.69 \\
1.44 & 20948 & 6450 & $ 1.63^{+ 0.03}_{- 0.04}$ & $ 0.84^{+ 0.04}_{- 0.04}$ & $ 1.95^{+ 0.08}_{- 0.08}$ & $ 34.14^{+18.51}_{- 8.88}$ & 0.62 & 0.14 & 0.55 \\
1.52 &  7396 & 2383 & $ 1.79^{+ 0.05}_{- 0.06}$ & $ 0.78^{+ 0.08}_{- 0.09}$ & $ 1.64^{+ 0.12}_{- 0.12}$ & $ 12.65^{+ 3.12}_{- 2.11}$ & 0.66 & 0.13 & 0.87 \\
1.65 &  1856 &  634 & $ 2.01^{+ 0.07}_{- 0.06}$ & $ 0.99^{+ 0.12}_{- 0.10}$ & $ 1.71^{+ 0.20}_{- 0.16}$ & $ 13.17^{+ 6.91}_{- 2.88}$ & 0.85 & 0.23 & 0.76 \\
1.74 &  2640 &  874 & $ 2.04^{+ 0.06}_{- 0.05}$ & $ 0.82^{+ 0.10}_{- 0.11}$ & $ 1.70^{+ 0.17}_{- 0.16}$ & $ 14.31^{+ 6.61}_{- 3.30}$ & 0.91 & 0.24 & 0.68 \\
1.85 &  2829 &  951 & $ 2.27^{+ 0.11}_{- 0.09}$ & $ 0.74^{+ 0.12}_{- 0.09}$ & $ 1.45^{+ 0.18}_{- 0.13}$ & $  8.90^{+ 2.56}_{- 1.35}$ & 0.95 & 0.25 & 0.71 \\
1.95 &  3030 &  961 & $ 2.24^{+ 0.13}_{- 0.07}$ & $ 0.78^{+ 0.09}_{- 0.11}$ & $ 1.37^{+ 0.13}_{- 0.17}$ & $  8.92^{+ 1.69}_{- 1.56}$ & 1.00 & 0.24 & 0.79 \\
2.79\tablenotemark{g} &  2802 &  928 & $ 2.46^{+ 0.39}_{- 0.09}$ & $ 0.65^{+ 0.08}_{- 0.04}$ & $ 1.30^{+ 0.11}_{- 0.10}$ & $  9.08^{+ 1.39}_{- 0.69}$ & 1.44 & 0.23 & 0.68 \\

\enddata
\tablenotetext{a}{Blackbody temperature}
\tablenotetext{b}{Blackbody radius}
\tablenotetext{c}{Cut-off power law photon index}
\tablenotetext{d}{Cut-off power law cut-off energy or lower limit}
\tablenotetext{e}{Total luminosity, $\times 10^{37}$ ergs 
s$^{-1}$, for a distance of 9.3 kpc, 2--50 keV}
\tablenotetext{f}{Ratio of blackbody flux to power law flux, 2--50
keV}
\tablenotetext{g}{Included gaussian line; $E_l = 6.14^{+0.18}_{-1.07}$
keV}
\tablecomments{For all fits $N_H$ is frozen at $0.2 \times 10^{22}$
cm$^{-2}$.  Errors are $1\sigma$ for one interesting parameter.}
\end{deluxetable}


\clearpage

\begin{deluxetable}{lcccccccccc}
\tablecaption{Spectral Fits of 4U 1820-30 with the CompTT + BB
Model\label{tab-fits2}} 
\tablewidth{0pt}
\tablehead{
\colhead{$S_a$} & 
\colhead{$kT_{BB}$} &
\colhead{$R_{BB}$} &
\colhead{$\tau$\tablenotemark{a}} &
\colhead{$kT_e$\tablenotemark{b}} &
\colhead{$kT_W$\tablenotemark{c}} &
\colhead{$y$\tablenotemark{d}} &
\colhead{$R_W$\tablenotemark{e}} &
\colhead{$L$} &
\colhead{$L_{BB}/L_H$} &
\colhead{$\chi^2_{\nu}$} \\
\colhead{} &
\colhead{(keV)} &
\colhead{(km)} &
\colhead{} & 
\colhead{(keV)} &
\colhead{(keV)} &
\colhead{} & 
\colhead{(km)} &
\colhead{} &
\colhead{} &
\colhead{}
}
\startdata
1.36 & $ 1.49^{+ 0.08}_{- 0.04}$ & $ 0.95^{+ 0.07}_{- 0.10}$ & $ 7.63^{+ 0.51}_{- 1.48}$ & $ 6.80^{+ 2.77}_{- 0.62}$ & $ 0.37^{+ 0.06}_{- 0.09}$ & 3.10 & $20.54^{+ 7.38}_{- 9.94}$ & 0.52 & 0.16 & 0.68 \\
1.44 & $ 1.63^{+ 0.03}_{- 0.06}$ & $ 0.89^{+ 0.08}_{- 0.04}$ & $ 6.40^{+ 1.16}_{- 1.00}$ & $ 8.34^{+ 2.55}_{- 1.87}$ & $ 0.39^{+ 0.04}_{- 0.07}$ & 2.67 & $20.92^{+ 5.33}_{- 8.48}$ & 0.61 & 0.19 & 0.54 \\
1.52 & $ 1.72^{+ 0.11}_{- 0.07}$ & $ 1.01^{+ 0.04}_{- 0.10}$ & $ 8.74^{+ 0.69}_{- 1.80}$ & $ 4.96^{+ 1.78}_{- 0.43}$ & $ 0.46^{+ 0.04}_{- 0.03}$ & 2.97 & $15.13^{+ 3.50}_{- 3.32}$ & 0.65 & 0.21 & 0.84 \\
1.65 & $ 1.93^{+ 0.07}_{- 0.09}$ & $ 1.15^{+ 0.09}_{- 0.08}$ & $ 8.08^{+ 1.86}_{- 3.55}$ & $ 5.34^{+ 8.39}_{- 1.16}$ & $ 0.46^{+ 0.04}_{- 0.06}$ & 2.73 & $16.76^{+10.50}_{- 6.89}$ & 0.84 & 0.36 & 0.74 \\
1.74 & $ 2.03^{+ 0.04}_{- 0.07}$ & $ 0.90^{+ 0.09}_{- 0.09}$ & $ 5.78^{+ 2.71}_{- 4.24}$ & $ 8.85^{+67.12}_{- 3.67}$ & $ 0.50^{+ 0.03}_{- 0.04}$ & 2.31 & $15.53^{+41.47}_{- 8.65}$ & 0.92 & 0.43 & 0.63 \\
1.85 & $ 1.87^{+ 0.12}_{- 0.18}$ & $ 0.91^{+ 0.10}_{- 0.10}$ & $12.43^{+ 1.23}_{- 1.53}$ & $ 3.44^{+ 0.43}_{- 0.26}$ & $ 0.39^{+ 0.06}_{- 0.21}$ & 4.16 & $21.81^{+ 6.74}_{-23.69}$ & 0.93 & 0.19 & 0.67 \\
1.95 & $ 2.03^{+ 0.12}_{- 0.15}$ & $ 1.00^{+ 0.08}_{- 0.08}$ & $10.49^{+ 1.97}_{- 3.10}$ & $ 4.09^{+ 2.11}_{- 0.58}$ & $ 0.50^{+ 0.05}_{- 0.05}$ & 3.52 & $13.95^{+ 4.41}_{- 4.28}$ & 0.99 & 0.34 & 0.78 \\
2.79 & $ 1.75^{+ 0.13}_{- 0.14}$ & $ 1.00^{+ 0.11}_{- 0.08}$ & $13.46^{+ 0.74}_{- 0.89}$ & $ 3.53^{+ 0.22}_{- 0.15}$ & $ 0.34^{+ 0.07}_{- 0.34}$ & 5.01 & $33.71^{+14.74}_{-67.25}$ & 1.41 & 0.15 & 0.71 \\

\enddata
\tablenotetext{a}{Comptonizing cloud optical depth}
\tablenotetext{b}{Comptonizing electron temperature}
\tablenotetext{c}{Temperature of seed photons (assumed to follow a
Wien law)}
\tablenotetext{d}{Comptonization parameter $y = 4kT_e\tau^2/m_ec^2$}
\tablenotetext{e}{Equivalent Wien radius of seed photons (see text)}
\tablecomments{For all fits $N_H$ is frozen at $0.2 \times 10^{22}$
cm$^{-2}$.  Errors are $1\sigma$ for one interesting parameter.} 
\end{deluxetable}


\clearpage

\begin{figure}
\figurenum{\ref{fig-asm}}
\plotone{asm_lc_paper.eps}
\caption{}
\end{figure}

\clearpage
 
\begin{figure}
\figurenum{\ref{fig-ccd}}
\plotone{ccd_paper.eps}
\caption{}
\end{figure}

\clearpage

\begin{figure}
\figurenum{\ref{fig-twospec}}
\plotone{twospec_paper.eps}
\caption{}
\end{figure}

\clearpage

\begin{figure}
\figurenum{\ref{fig-fit}}
\plotone{fit_spec.eps}
\caption{}
\end{figure}

\clearpage

\begin{figure}
\figurenum{\ref{fig-params1}}
\plotone{bb_cpl_params_paper.eps}
\caption{}
\end{figure}

\clearpage

\begin{figure}
\figurenum{\ref{fig-params2}}
\plotone{bb_cs_params_paper.eps}
\caption{}
\end{figure}

\clearpage

\begin{figure}
\figurenum{\ref{fig-comp}}
\plotone{comp_paper.eps}
\caption{}
\end{figure}

\clearpage

\begin{figure}
\figurenum{\ref{fig-ccds}}
\plotone{ccds_together.eps}
\caption{}
\end{figure}

\end{document}